# Laser patterned diamond electrodes for adhesion and proliferation of human mesenchymal stem cells

Hassan N. Al Hashem[1], Amanda N. Abraham[2], Deepak Sharma[1,3,4], Andre Chambers[5], Mehrnoosh Moghaddar[2], Chayla L. Reeves[2], Sanjay K. Srivastava[3,4], Amy Gelmi[2*], and Arman Ahnood[1**]

[1]School of Engineering, RMIT University, Melbourne, Victoria 3000, Australia

[2]School of Science, RMIT University, Melbourne, Victoria 3000, Australia

[3]Photovoltaic Metrology Section, Advanced Materials and Device Metrology Division, CSIR-National Physical Laboratory, Dr K. S. Krishnan Marg, New Delhi-110012, India

[4]Academy of Scientific and Innovative Research (AcSIR), Ghaziabad-201002, India

[5]School of Engineering, University of Melbourne, Melbourne, Victoria 3010, Australia

\* Corresponding author: amy.gelmi@rmit.edu.au
\*\* Corresponding author: arman.ahnood@rmit.edu.au

Keywords: Diamond electrodes, Stem cells, Diamond substrate, graphitization, Capacitance, Impedance, oxygen termination, nitrogen termination.

## ABSTRACT

The ability to form diamond electrodes on insulating polycrystalline diamond substrates using single-step laser patterning, and the use of the electrodes as a substrate that supports the adhesion and proliferation of human mesenchymal stem cells (hMSCs) is demonstrated. Laser induced graphitisation results in a conductive amorphous carbon surface, rich in oxygen and nitrogen terminations. This results in an electrode with a high specific capacitance of 182 $\mu$F/cm$^2$, a wide water window of 3.25 V, and a low electrochemical impedance of 129 $\Omega\cdot$cm$^2$ at 1 kHz. The electrode's surface exhibited a good level of biocompatibility with hMSCs, supporting cell adhesion and proliferation. The cells cultured on the electrode displayed significant elongation and alignment along the direction of the laser patterned microgrooves. Because of its favourable electrochemical performance and biocompatibility, the laser-patterned diamond electrodes create a potential for a versatile platform in stem cell therapeutics.

# 1. INTRODUCTION

As a bioinert material with excellent electrochemical attributes, diamond-based electrodes have been used for a number of bio-interfacing applications to modulate and monitor cellular activities. Examples of this include multimodal electrodes for neural stimulation [1, 2] and recording [3, 4], biomolecular [5] and DNA [6] sensing, as well as monitoring of cardiac action potentials [7]. Specifically in these applications, the low electrochemical impedance and high charge injection capacity of diamond-based electrodes [8, 9] make them particularly effective for bio-interfacing. Diamond has also been used as a substrate to support the development of cellular networks such as cortical neurons [8, 10], human neural stem cells [11], and sperm cells [12]. Here, the versatile surface chemistry, morphology and hardness of the diamond substrate have facilitated the spontaneous differentiation of human neural stem cells [13], the osteoblastic differentiation of human mesenchymal stem cells [14] or optical control of neuronal development [15] amongst others. Although earlier works have demonstrated the potential for electrical bio-interfacing with stem cells [16-20], diamond electrodes have not yet been extensively investigated for this purpose. Earlier works include boron doped diamond [11] and nitrogen doped ultrananocrystalline diamond [21]

A key challenge in developing diamond electrodes for bio-interfacing has been its manufacturability. While scalable diamond substrate production has been possible and is in use for a myriad of commercial applications, diamond in its pure form is an insulator. The primary approach for creating conductive diamond electrodes has centred on the incorporation of dopants during their deposition [22, 23]. Although this approach has fuelled much of the research in diamond-based electrodes, it has prevented patterned electrode fabrication using commercially produced insulating diamonds, which are being grown at industrial scales. Direct modification of an insulating diamond to conductive structures after its growth is possible. Ion-beam [24-26], and laser [27-29] graphitisation have been demonstrated to create heterogeneously integrated conductor/insulator structures.

This work reports a diamond-based electrode fabricated using laser-induced graphitisation of a commercially available insulating diamond substrate and its viability to support the development of human mesenchymal stem cells. An ultraviolet nanosecond laser is used for single-step direct writing of electrode structure on insulating polycrystalline diamond (PCD). Surface chemistry, structure and morphology of the newly developed electrode material are reported and discussed in the context of other diamond-based materials. The key electrochemical attributes of the laser-patterned PCD electrodes are presented and compared to those of conventional bio-interfacing electrodes. The growth of human mesenchymal stem cells (hMSCs) on the laser-patterned PCD electrode over 72-hour window is assessed, and its relation to the morphological features of the electrodes is evaluated.

# 2. METHODS

## 2.1 Laser Micromachining

The diamond substrate used in laser writing is 10 mm × 10 mm × 0.3 mm thermal management grade PCD purchased from Element Six (TM100). The whole PCD surface was laser patterned using an ultraviolet (UV) nanosecond pulsed laser micromachining system (DCT, DL561U) at ambient atmosphere and temperature. The employed UV laser has a wavelength of 355 nm, average power of ~4 W, pulse width of ~16 ns, repetition rate of 30 kHz, speed of 200 mm/s, 2 laser passes, and 133 µJ/pulse. The laser ablation was performed in parallel lines with a line interval of 5 µm and 50% overlap.

## 2.2 Material Characterisation

The morphology of the electrode surface was examined using Scanning Electron Microscopy (SEM) with FEI Quanta 200 at 30 kV at different magnifications. The surface topography was assessed using three-dimensional optical profilometry (Bruker ContourGT-K) with illumination wavelength of 546 nm and through a 20X objective lens, covering an area of 320 µm × 240 µm. Gwyddion was used to analyse and visualise the profilometry data. The surface topography was measured at the centre of the laser-patterned PCD.

Raman spectroscopy was acquired at room temperature using a Raman microscope (LabRAM HR Evolution) at excitation wavelength of 532 nm, 1.6 mW, 100X objective lens, and 1800 gr/mm grating. To estimate the ratio of D- and G-band peak intensities I(D)/I(G), the spectra background was subtracted using the straight-line method. OriginPro software was used in background subtraction, normalisation, and visualisation of Raman spectroscopy data.

X-ray photoelectron spectroscopy (XPS) was acquired using Kratos Axis Supra using Aluminum Ka X-rays as a source gun with $E_{photon}$ of 1486.7 eV. CasaXPS software was used in the quantification of elements, removing the background by utilising "U 2 Tougaard" model [30]. The sheet conductivity was approximated by measuring the sheet resistance using a four-point probe (Janedel RM3000) and calculating the resistivity of the graphitic layer [31], assuming a thickness of 1 µm.

## 2.3 Electrochemical Characterisation

The cyclic voltammetry and electrochemical impedance spectroscopy (EIS) measurements were recorded using CHI 760D potentiostat (CH Instruments Inc.), while the multi-potential and multi-current steps were recorded using CHI 920C potentiostat (CH Instruments Inc.). The electrochemical cell comprises a chamber, the sample (laser-patterned PCD), and a typical 0.15 mol/L NaCl (saline) solution. This concentration was used to mimic the biological environment in line with our earlier work [32]. The chamber was fabricated using a three-dimensional printer with acrylonitrile butadiene styrene (ABS) filament, and subsequently coated with a two-part epoxy resin. A three-electrode configuration was used in this setup. Silver/Silver Chloride (Ag/AgCl) (eDAQ-ET054) was used as reference electrode, 0.5 mm Platinum (Pt) as counter electrode, and the sample as working electrode. The sample area exposed to the saline solution is circular with a radius of approximately 2.7 mm resulting in an area of 0.23 cm$^2$. The electrochemical double layer capacitance value was obtained through a linear fitting on the plot of voltage scan rates versus the current at 0 V based on the capacitance equation: $i_C = C \frac{dV}{dt}$, where $i_C$ is the capacitive current in µA, $C$ is capacitance in µF, and $dV/dt$ represent the scan rate in V/s. The utilised scan rates ranged between 10 – 250 mV/s. In order to determine the electrochemical water window of the electrode, multiple CV measurements were conducted at a broad range of voltage limits from ± 0.5 V to ± 2.5 V. Subsequently, the region exhibiting a lack of exponential changes in current was used as the water window. The EIS was conducted at 10 mV amplitude, D.C. offset of zero, relative to the reference electrode with a range of frequencies from 1 Hz to 500 kHz.

In the experiments of both multi-potential and multi-current steps, a sampling interval of 0.001 s was used. The current was recorded with a sample sensitivity of 0.001 A/V in the multi-potential steps experiment. Due to the limitations of the potentiostat of being unable to record continuous data for extended periods of time, multipotential and multicurrent steps were recorded in groups. Then, a Python program was developed to stitch the data over the course of 12 hours, and afterwards, extract representative readings in intervals of 10 minutes. The positive and negative peaks were detected by finding the local maxima and minima via SciPy signal processing library [33]. Pandas [34, 35] and NumPy [36] libraries were used in the basic operations like reading, concatenating and arranging data. The rates of change in current and potential were plotted as $\frac{I-I_0}{I_0}$, and $\frac{V-V_0}{V_0}$ respectively, where $I$ is the current, $I_0$ is the initial current, $V$ is the potential, and $V_0$ is the initial potential. All electrochemical stability data were analysed and visualised using OriginPro.

## 2.4 Cell Culture of hMSCs

Bone marrow-derived human mesenchymal stem cells (hMSCs) (Lonza) were initially revived at passage 4 and cultured in growth media (Mesenchymal Stem Cell Growth Medium, Lonza USA) at 37°C and 5% $CO_2$. After reaching 80-90% confluency, the hMSCs were seeded at a cell density of $2 \times 10^5$ cells/mL was seeded onto the substrates (PCD and tissue culture plate (TCP)). The hMSCs were maintained in basal (Minimum Essential Medium Alpha (α-MEM), 10% FBS, 1% pen-strep) media for the duration of the experiments, and media was changed every 3-4 days.

## 2.5 Cell Viability of hMSCs

A resazurin assay was used to investigate cell viability by monitoring cellular metabolic activity over time. The assay was performed at 24, 48, and 72 hours post-seeding. The assay solution was comprised of 0.01 mg/mL resazurin sodium salt (Sigma-Aldrich, R7017) in a phenol-free α-MEM, and the seeded cells were incubated in the assay solution for two hours at 37°C and 5% $CO_2$. The supernatant was then collected and fresh basal media replenished. The fluorescence intensity of the supernatant aliquots (n=3) was measured using a plate reader (ClarioStar) with a 545 nm excitation wavelength. The statistical analysis of fluorescence data was performed and displayed using the software GraphPad Prism.

## 2.6 Cell Fixing and Staining

The cells seeded on the PCD substrates were fixed after 3 days of cell culture with 4% paraformaldehyde (PFA), then triple washed and stored in phosphate buffered saline (PBS) until staining. The cells were prepared for immunofluorescent staining with permeabilization using 0.25% Triton X-100 (Thermo Fisher Scientific Inc.) in PBS, triple washed in PBS, and blocked in 1% bovine serum albumin (BSA) (Sigma-Aldrich, A2934) in PBS for 30 minutes at room temperature, then triple washed in PBS. The cells were then incubated in the primary antibodies for vinculin (Sigma-Aldrich, V9131) at 1:400 in PBS overnight at 4°C and triple washed in PBS. The secondary antibody of donkey anti-mouse (Alexa Fluor® 555, Abcam-ab150106) diluted in PBS to 1:500, Phalloidin-iFluor 488 (ab176753) for staining F-actin filaments, and NucBlue (Invitrogen™, R37605) for staining nuclei, were added and the cells were incubated at room temperature for 45 minutes. The cells were then triple washed in PBS and stored in PBS at 4°C until imaging.

## 2.7 Cell Imaging and Alignment Analysis

The immunofluorescent imaging of the cells on laser-patterned PCD were conducted using OLYMPUS IX83 with excitation wavelengths of 390 nm, 485 nm, and 560 nm at different magnifications. The images of each wavelength were merged and displayed using Fiji software [37].

To assess cellular alignment, F-actin images were analysed using the Directionality plugin [38] (version 2.3.0) in Fiji, which performs Fourier spectrum analysis across a range of 0° to 180°. Image brightness and contrast were optimised, and a background subtraction was performed in Fiji to minimise the background involvement in the analysis process.

## 3. RESULTS AND DISCUSSION

### 3.1 Surface chemistry and morphology

The Scanning Electron Microscopy (SEM) images of the laser patterned diamond electrodes in Figure 1(a), (b) and (c) show the surface morphology of the laser micromachined PCD. The laser ablation of the PCD surface results in the appearance of a repetitive pattern, as shown in Figure 1(a). A series of distinct round cavities of ∼5 µm in diameter can be observed throughout the surface, which seems to be caused by the nanosecond laser pulses. The pattern generated by laser pulses mainly exhibits linearity. However, in certain regions, the presence of round cavities formed by these pulses becomes more pronounced, leading to a dotted-line appearance. Figure 1(b) a close-up view of one of the cavities shows the porous nature of a typical sidewall surface morphology. Figure 1(c), the side-wall of the cavities exhibits a distinct ripple-like shape, which is a phenomenon that is reported previously and linked to the wavelength utilised in a linearly polarised laser and its laser incident angle [39]. Figure 1(d) shows an array of laser patterned diamond electrodes with dimensions of 250 µm × 250 µm and a pitch size of 250 µm.

Figures 1(e) and (f) show the contrast in the surface chemistry between the surfaces of the laser patterned electrode areas and the insulating PCD substrate. X-ray photoelectron spectroscopy (XPS) survey depicted in Figure 1(e) investigates the surface elemental states. The XPS survey of the PCD contained mostly carbon and traces of oxygen, as detailed in Table 1. Compared to the pristine PCD, the laser-patterned PCD displayed a noticeable introduction of nitrogen onto the PCD surface, alongside a significant increase in oxygen content relative to the carbon content. These variations suggest the absorption of ambient gases during the laser writing process. Similar observations were reported in earlier works on laser annealing experiments, where various species were transfer doped from gas phase onto the surface of diamond-like carbon and polycrystalline diamond [40], and single-crystalline diamond [39].

.

| **Chemical Composition** | **Insulating PCD substrate** | **Laser patterned electrode** |
|---|---|---|
| **C** (%) | 95.24 | 65.66 |
| **O** (%) | 4.76 | 22.47 |
| **N** (%) | 0.0 | 11.87 |

Table 1 Atomic percentages of the contributing elements acquired using XPS, highlighting significant changes in carbon, oxygen and nitrogen content before and after laser patterning.

Raman spectroscopy was used to study the chemical composition and crystallinity of PCD substrate and laser patterned diamond electrodes, as shown in Figure 1(f). In the PCD substrate, a distinct peak is visible at 1332 cm$^{-1}$, which is attributed to the first-order diamond Raman line [41]. As opposed to this, the spectra of laser patterned electrode shows broad peaks at 1347 cm$^{-1}$ and 1587 cm$^{-1}$, which corresponds to the D-band and G-band, respectively [41]. The transition from the first-order diamond Raman line to the emergence of D-band and G-band peaks suggests graphitisation of the PCD surface induced by the laser writing process. The ratio of D- and G-band peak intensities, denoted as I(D)/I(G), was estimated to be 0.833. As highlighted in Table 2, the laser patterned PCD exhibits lower I(D)/I(G) ratio compared to alternative diamond electrodes with graphitic carbon inclusion, such as nitrogen-doped ultrananocrystalline diamond (N-UNCD) [9, 42], suggesting a more ordered sp$^2$ hybridised structure. Unlike N-UNCD, the absence of transpolyacetylene (TPA) segments Raman lines at 1140 cm$^{-1}$ and 1480 cm$^{-1}$ [9, 43, 44] after laser patterning indicates minimal hydrogen incorporation from the ambient atmosphere.

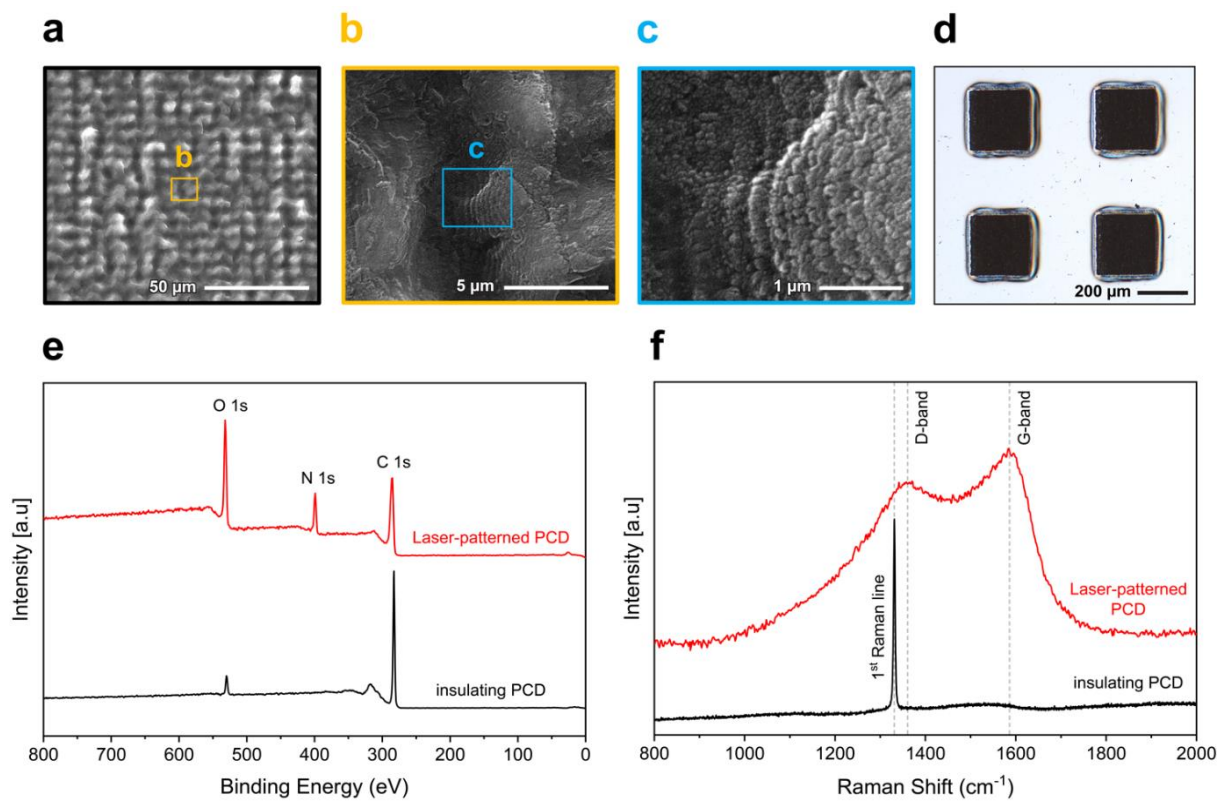

Figure 1 (a), (b) and (c) SEM images of the laser patterned diamond electrodes at different magnifications showing the morphological features of the electrode. (d) Optical microscope image of an array of four laser patterned diamond electrodes, dark regions with dimensions of 250×250 μm, on an insulating PCD substrate. (e) XPS survey data of the laser patterned diamond electrode compared to the insulating PCD substrate. (f) Raman spectra of laser patterned diamond electrode compared to the PCD substrate.

To explore the surface characteristics of the laser patterned PCD, three-dimensional optical profilometry was performed. The surface features exhibited heights ranging approximately from -5 μm to 5 μm, with a root mean square roughness (Sq) estimated at 1.21 μm, as opposed to <1 nm of pristine PCD [45]. Upon comparing the results from three-dimensional optical profilometry with SEM images, a consistent pattern of laser pulses can be observed in both images, as shown in Figure S1 in supporting information.

| Material | I(D) / I(G) | Conductivity (S/cm) | Ref |
|---|---|---|---|
| As-grown N-UNCD | 1.191 | 46 | [9, 42] |
| Oxygen-annealed N-UNCD | 1.101 | – | [9] |
| Laser-patterned electrode | 0.833 | 34.30 | This work |

Table 2 The ratio of D- and G-band peak intensities of Raman spectroscopy, denoted as I(D)/I(G), and sheet conductivity of the graphitic layer.

### 3.2 Electrochemical Assessment

In Figure 2(a), cyclic voltammetry (CV) was performed between −0.5 V and 0.5 V to cover a wide range and avoid most redox currents [46]. The CV measurement was repeated at various voltage scan rates from 10 mV/s to 250 mV/s. The correlation between scan rates and the current at 0 V exhibited a linear relationship with a high coefficient of determination ($R^2$= 0.994), as displayed in Figure 2(b). The derived double layer capacitance value was approximately 41.9 µF. Upon normalisation by the electrode area (0.23 cm$^2$), this yields an effective double layer capacitance of 182 µF/cm$^2$. It is noteworthy the recorded current within the -0.5 V to 0.5 V lacks exponential changes and exhibits a more rectangular shape. This characteristic signifies a capacitive behaviour, wherein the current is passing indirectly via redistributing the charges in the electrolyte [47]. The laser patterned PCD exhibits relatively high double-layer capacitance when compared to most metals and carbon-related electrodes, as shown in Table 3. This facilitates a greater capacity of charge injection, thus offering significant advantages for stimulation purposes. Indeed, oxygen surface functionalisation, as seen in oxygen terminated diamond electrodes such as N-UNCD, can lead a substantial change in magnitude of double-layer capacitance [9].

In Figure 2(c), the water window was estimated to be 3.25 V through conducting a series of CV measurements across different voltage ranges at a fixed scan rate of 100 mV/s, and subsequently identifying the region exhibiting minimal sharp changes. Characterising the water window is a crucial for electrodes, as it signifies a safe operating region for biomedical applications [48]. As illustrated in Table 3, laser patterned PCD exhibits a notably wide water window in comparison to other metallic and carbon-related electrodes. It should be noted that the +/- 2.5 V trace in Figure 2 (c) exhibits a redox peak at +0.5 V, which can be attributed to graphitic materials. This peak is absent when narrower voltage ranges are used, suggesting that it relates to surface changes due to the use of voltages beyond the water window.

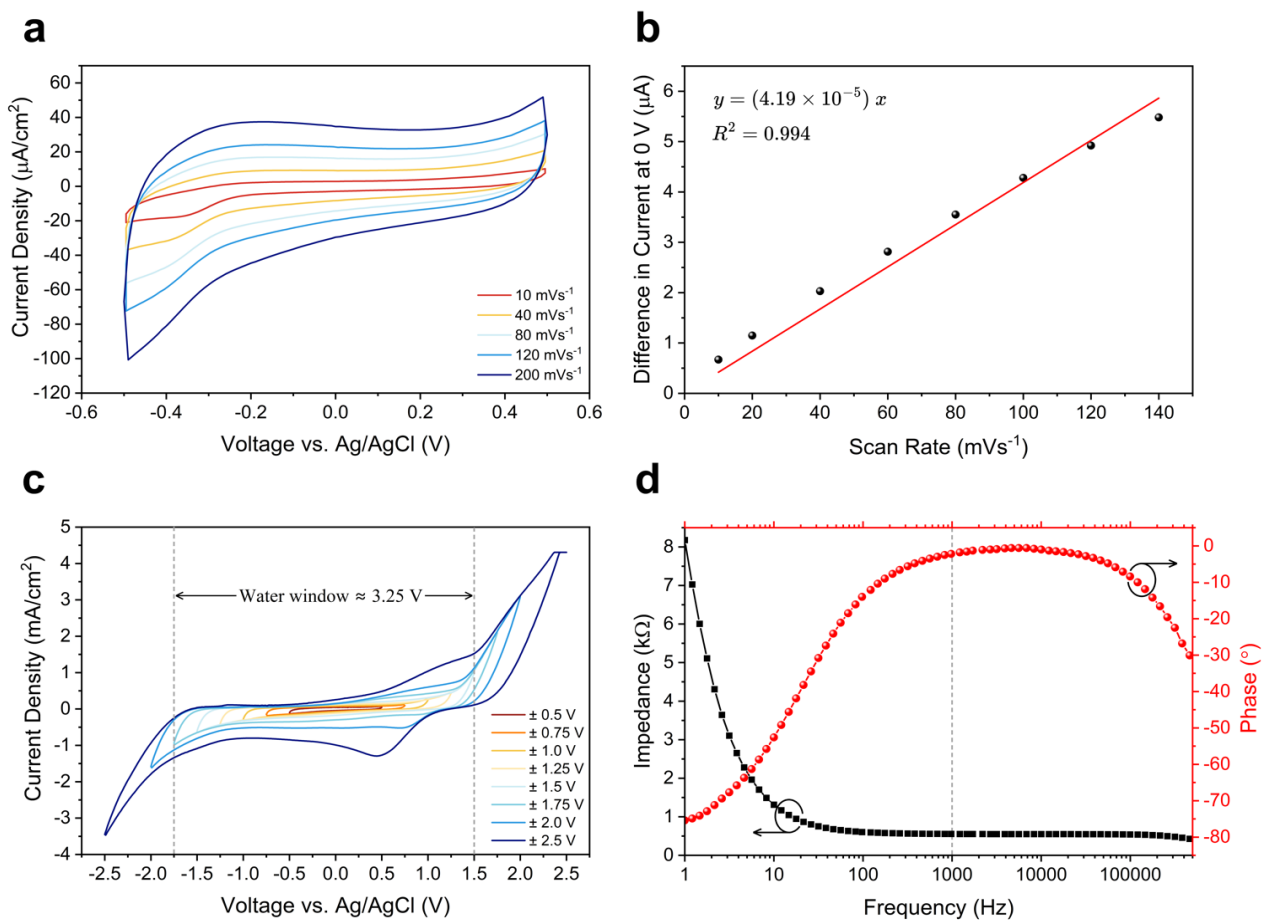

Figure 2 (a) Cyclic voltammetry sweeps between -0.5 to 0.5 V at various scan rates ranging from 10 mV/s to 250 mV/s. (b) the relationship between scan rates versus difference in current at 0 V from CV measurement. (c) Multiple CV measurements conducted at fixed scan rate of 100 mV/s across varying voltage ranges from ± 0.5 to ± 2.5 V, aimed at evaluating the electrochemical water window. (d) Electrochemical impedance spectroscopy of the laser patterned PCD, showing the impedance magnitude and phase at frequencies ranging from 1 to 500,000 Hz.

In Figure 2(d), EIS measurement was conducted over a frequency range from 1 Hz to 500 kHz to characterise the electrochemical impedance of laser patterned PCD. The impedance magnitude at 1 kHz was found to be 554 Ω, corresponding to a specific electrochemical impedance of 129 Ω·cm$^2$. At this frequency, the phase angle is approximately -5°, suggesting that the impedance is dominated by the resistive contributions of the electrolyte and a minimal electrode contribution. This compares favourably to conventional platinum electrodes, which exhibit a much higher phase angle, indicating the dominance of the capacitive electrode impedance - for example -75° at 1 kHz reported by [49]. Notably, the electrochemical impedance remained relatively stable over a wide frequency range from approximately 100 to 100,000 Hz, and the phase angle approaches zero degrees at relatively low frequencies – which is consistent with other diamond-based electrodes with oxygen-rich surfaces [9]. Compared to commonly used metal and carbon-based electrodes, the impedance magnitude of laser-patterned PCD is notably higher, as demonstrated in Table 3. Low electrochemical impedance is desirable for stimulating electrodes, facilitating higher charge injection without excessive power consumption [50].

| Material | Specific double-layer capacitance (µF/cm$^2$) | Ref | Specific electrochemical impedance (Ω·cm$^2$) | Ref | Water Window (V) | Ref |
|---|---|---|---|---|---|---|
| Oxygen annealed N-UNCD | 3746 ± 132 | [9] | 14 | [9] | 3.1* | [8] |
| Titanium nitride | 1050 – 1275 | [51] | – | – | 1.8 | [52] |
| Platinum | 49 | [53] | 4.9 | [54] | 1.4 | [55] |
| As-grown N-UNCD | 38 | [42] | 46 | [42] | 3.1 | [42] |
| Graphene | 21 | [56] | 13.5 | [57] | 1.2 | [58] |
| Gold | 14 | [59] | 4.5 | [60] | 1.4 | [55] |
| Boron-doped diamond | <10 | [61] | ~140 | [62] | 3 | [63] |
| Heavily Boron-doped diamond (B/C=5%) | 214 | [64] | – | – | 1.6 | [64] |
| Laser-patterned PCD | 182 | This work | 129 | This work | 3.25 | This work |

Table 3 A comparison of specific double-layer capacitance, specific electrochemical impedance, and water window of popular metal and carbon-based electrodes. *Oxygen terminated N-UNCD.

To investigate the stability of the electrode during stimulation in terms of maintaining a stable current amplitude and for an extended period of time, two experiments were conducted over 12 hours. In the first experiment, a cyclic pattern of current pulses was generated while the potential is recorded, via a multiple-current steps technique, as shown in Figure 3. The current profile of an individual cycle begins with a rest period of 10 s, followed by a positive and negative current pulse of 0.1 mA for 0.05 s, with a rest period of 0.45 s between the two pulses, as shown in Figure 3(a). Afterwards, the cycle was repeated for approximately 12 hours. In the interest of simplicity, the graph shows one cycle every 10 minutes, while displaying only the highest point, called peak, and the lowest point, called negative peak, as shown in Figure 3(b). The stability of the potential response over the 12 hours was evaluated utilising linear fitting, as illustrated in Figure S2 in supporting information. The average rate of potential change was found to be around 4.74 mV/1000 pulses and 1.30 mV/1000 pulses for the positive and negative peaks, corresponding to 0.95 %/1000 pulses and 1.71 %/1000 pulses, respectively. Here, both positive and negative peaks drifted in the same direction, which may indicate charge buildup on the electrode surface due to the stimulation configuration. In the second experiment, a multiple-potential steps technique was used to generate a cyclic pattern of potential pulses while recording the current, as shown in Figure 9. The potential profile for each cycle starts and ends with a rest period of 0.225 s, and in between, there is a positive and negative potential pulse of 0.1 V for 0.05 s, and a rest period between the two pulses for 0.45 s, as illustrated in Figure 3(c). Then, the cycle was run on a loop for about 12 hours. Similarly to the first experiment, only one cycle every 10 minutes is shown in Figure 3(d), which is simplified by displaying only the positive and negative peaks. Over the 12 hours, the average rates of current change for the positive and negative peak amplitudes are approximately -0.042 µA/1000 pulses (-0.021 %/1000 pulses) and 0.037 µA/1000 pulses (0.018 %/1000 pulses), respectively. This indicates a reduction in the peak-to-peak current over the stimulation cycle, which may be an indicator of change in the electrode surface. The minimal variations in current and potential responses in both experiments suggest high stability of the electrode with only a slight change in potential and current over long period. This underscores the advantage of delivering a consistent potential amplitude and charge injection, especially when used in stimulation applications. Additionally, it suggests that the electrode maintains durability over prolonged usage.

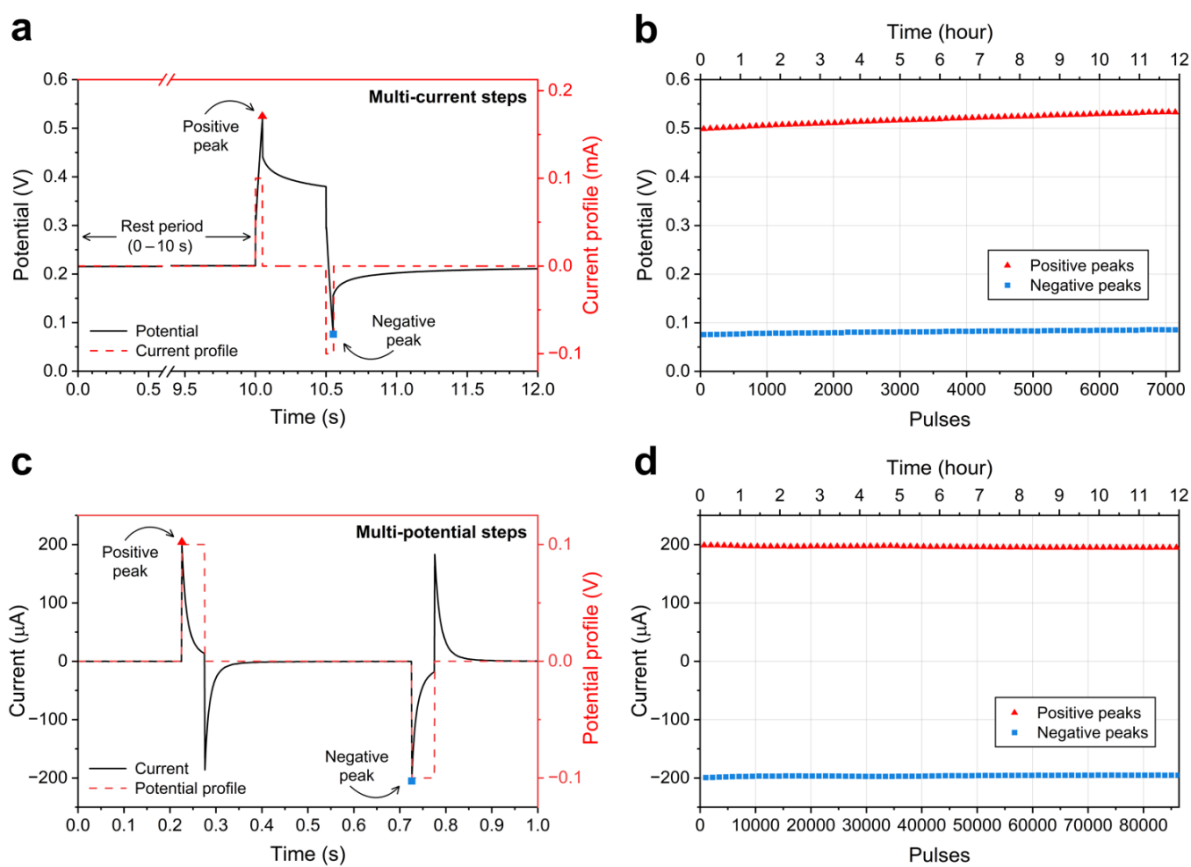

Figure 3 (a) A cyclic pulse of the applied potential profile and the recorded current for 12 seconds. (b) graph of potential positive and negative peaks over 12 hours utilised using multi-current steps technique. (c) A sample of the cyclic pulse showing the applied potential profile and the recorded current as a function of time. (d) plot of current positive and negative peaks over 12 hours using multi-potential steps technique.

3.3 Biocompatibility of hMSCs

Merged immunofluorescence widefield images of hMSCs seeded on TCP control substrate is shown in Figure 4(a), whilst Figure 4(b) and (c) depict laser-patterned PCD substrate. They show merged images related to different intracellular components of the stem cells; the actin (green) which is the main component of the cytoskeletal structure of the cells, vinculin (red) which indicates focal adhesions, and the nucleus (blue) of the cell. As shown in Figure 4(b), the actin stain reveals that the cells grew to cover a substantial area over the laser patterned PCD, reaching more than 500 μm while branching and interconnecting with neighbour cells. The vinculin stain shows that the cell adhesion sites are mainly located near the centre of cells rather than the limbs. Additionally, the hMSCs cultured on laser patterned PCD exhibit a thin and elongated morphology aligned with the direction of the laser writing microgrooves as depicted in Figure 4(b). In contrast, considering that laser patterned PCD and TCP images have the same scale, it is clear that the cells grown on TCP more spread in comparison. Figure 4 (c) shows the higher magnification image, illustrating the cell conforming to square-like boundaries of the laser patterned PCD and highlighting a single cell aligning itself with the topography of the substrate. Additional immunofluorescence images of hMSCs on laser-patterned PCD are shown in Figure S3 in supporting information.

In Figure 4(d), the results of the directionality analysis suggest that the surface topography of laser patterned PCD has influenced the cellular behaviour, leading to elongation and alignment along the microgrooves created by laser writing. On the contrary, the lack of pronounced directionality to a specific angle is observed in cells cultured on TCP which is typical for this material. This phenomenon has been consistently observed in microgrooved substrates across various types of materials, including TCP [65], silicon substrates [66-69], biopolymers [70], collagen–fibroin [71], and carbon-based materials [72, 73]. Moreover, the elongation of the cell body has been a subject of discussion in numerous studies, often associated with guiding the differentiation of human mesenchymal stem cells (hMSCs) toward specific cell lineages, such as neuronal lineage [68, 72, 73] , osteogenic lineage [69, 71], adipocyte lineage [66], myogenic lineage [66, 70] and myocardial lineage [74].

In Figure 4(e), the viability was assessed at multiple time points (24, 48, and 72 hours) during cell culture for hMSCs grown on laser pattered PCD in comparison to tissue culture plastic (TCP). The normalised fluorescence intensity represents the cellular metabolism, which serves as a measure of the viability of hMSCs. The viability of hMSCs on laser-patterned PCD is not significantly different to the control surface of TCP, which suggests that the surface topography and material characteristics resulting from laser writing of PCD promote cellular adhesion and does not inhibit the growth of hMSCs. Similar findings have been reported in studies involving other carbon-based materials, including graphitic carbon films [75], graphene [76], and different types of diamond [11, 14, 77]. The growth and proliferation of hMSCs on laser patterned PCD may be attributed to the topographical features of the substrate, creating a favourable microenvironment and supporting the cellular proliferation [78, 79].

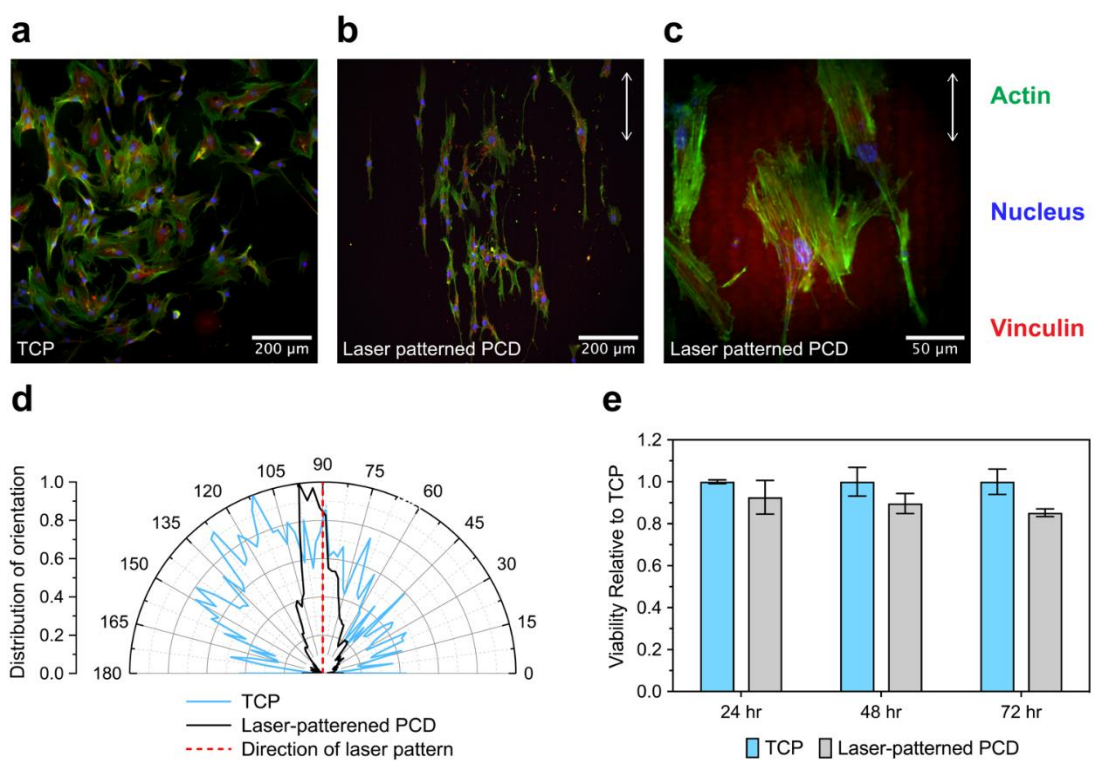

Figure 4 Merged immunofluorescence widefield images of hMSCs seeded on (a) TCP (control substrate) and (b) and (c) laser-patterned PCD. The images are acquired at different magnifications and wavelengths, where the green colour (485 nm) is the actin stain, the blue colour (390 nm) is the nuclear stain, and the red colour (560 nm) is the vinculin stain. The arrows represent the direction of laser writing microgrooves. (d) Alignment profile of hMSCs cultured on TCP and laser-patterned PCD after 3 days of growth, with alignment compared to the direction of laser writing microgrooves. (e) The resazurin assay, illustrating the normalised fluorescence intensity of hMSCs seeded on laser-patterned PCD relative to TCP after 24, 48, and 72 hours. The values are average ± standard deviation.

## 4. CONCLUSION

This work reports a single-step fabrication of diamond electrodes through direct graphitisation of polycrystalline diamond using nanosecond laser writing. The graphitisation process induced by the laser patterning resulted in the presence of highly-ordered sp$^2$ hybridised carbon. Moreover, the laser-induced graphitisation leads to the incorporation of oxygen and nitrogen in the surface of the electrode. Both attributes are akin to high-performing N-UNCD electrodes reported in our earlier works [8, 9]. The laser patterned electrodes exhibited a high specific double-layer capacitance of approximately 182 µF/cm$^2$ with a water window of 3.25 V. The electrochemical impedance was found to be about 129 Ω·cm$^2$ at 1 kHz. The electrochemical stability analysis, conducted over 12 hours, demonstrated a highly stable recorded potential and current, with a minimal average rate of change ranging between 0.037–4.74 %/1000 pulses, highlighting its suitability for long-term applications. The electrode's surface exhibited high biocompatibility with hMSCs in comparison to a conventional surface (TCP). Noticeably, the cells cultured on the electrode displayed significant elongation and alignment along the direction of the laser writing microgrooves, while on TCP, the cells lacked any observable elongation or alignment. Overall, this study demonstrates the potential of laser-patterned PCD electrodes for use in biological applications, benefiting from the favourable electrochemical performance and the biocompatible patterned surface. This work paves the way for further exploration of these electrodes in stem cell therapeutics and guiding the differentiation of stem cells into specialised cell types.

## ACKNOWLEDGEMENTS

The authors acknowledge the facilities, as well as the scientific and technical assistance of the RMIT University's Microscopy and Microanalysis Facility (RMMF), and the RMIT Micro Nano Research Facility (MNRF) in the Victorian Node of the Australian National Fabrication Facility (ANFF-Vic). H.N.A acknowledges the Ministry of Education in Saudi Arabia for funding a part of his Ph.D. scholarship and financial expenses.